\documentclass{new_tlp}
\usepackage[latin1]{inputenc}
\usepackage{mathptmx}
\usepackage{amsfonts,amssymb,latexsym}
\usepackage{amsmath}
\usepackage{algorithm}
\usepackage[noend]{algpseudocode}
\usepackage[all]{xy}
\usepackage{alltt,verbatim}
\usepackage{graphicx}
\usepackage{verbatim} 
\usepackage{url}
\usepackage[T1]{fontenc}
\usepackage{wrapfig}
\usepackage{caption}
\usepackage{multirow}

\usepackage{color} 
\definecolor{Blue}{rgb}{0.1,0.1,0.9}
\definecolor{Red}{rgb}{0.9,0.1,0.1}
\definecolor{Green}{rgb}{0.1,0.9,0.1}

\let\oldAA\AA
\renewcommand{\AA}{\text{\normalfont\oldAA}}

\def\defemb#1#2{\expandafter\def\csname #1\endcsname
	{\relax\ifmmode #2\else\hbox{$#2$}\fi}}

\def\res{\mathrel{\vert\grave{ }}}

\long\def\comment#1{}

\def\defemb#1#2{\expandafter\def\csname #1\endcsname{\relax\ifmmode #2\else\hbox{$#2$}\fi}}

\def\ll{[\![}
\def\rr{]\!]}
\def\Den#1{\relax\ifmmode \ll #1\rr \else\hbox{$\ll #1\rr$}\fi}
\def\res{\mathrel{\vert\grave{ }}}

\let\|=\mid

\def\#{\hat{~}}

\defemb{implies}{\quad\rightarrow\quad}
\defemb{Implies}{\quad\Rightarrow\quad}

\defemb{cA}{{\cal A}}
\defemb{cB}{{\cal B}}
\defemb{cC}{{\cal C}}
\defemb{cD}{{\cal D}}
\defemb{cE}{{\cal E}}
\defemb{cF}{{\cal F}}
\defemb{cG}{{\cal G}}
\defemb{cH}{{\cal H}}
\defemb{cI}{{\cal I}}
\defemb{cJ}{{\cal J}}
\defemb{cL}{{\cal L}}
\defemb{cP}{{\cal P}}
\defemb{cR}{{\cal R}}
\defemb{cS}{{\cal S}}
\defemb{cT}{{\cal T}}
\defemb{caT}{{\cal T}}
\defemb{cU}{{\cal U}}
\defemb{cV}{{\cal V}}
\defemb{cX}{{\cal X}}
\defemb{caX}{{\cal X}}
\defemb{cZ}{{\cal Z}}

\long\def\comment#1{}

\defemb{cR}{{\cal R}}
\defemb{cE}{{\cal E}}
\defemb{cN}{{\cal N}}
\defemb{cV}{{\cal V}}
\defemb{cP}{{\cal P}}
\defemb{cX}{{\cal X}}
\defemb{cU}{{\cal U}}

\newcommand\bcmdtab{\noindent\bgroup\tabcolsep=0pt
\begin{tabular}{@{}p{10pc}@{}p{20pc}@{}}}
\newcommand\ecmdtab{\end{tabular}\egroup}

\newcounter{@inst}
\newcounter{@auth}

\newdimen\instindent
\newbox\authrun
\newtoks\authorrunning
\newtoks\tocauthor
\newbox\titrun
\newtoks\titlerunning
\newtoks\toctitle
\newcommand{\nt}[1]{\mbox{$\langle$\textit{{#1}}$\rangle$}}
\def\Glints{\textup{\textsf{GLINTS}}}

\begin{document}
\title[Theory and Practice of Logic Programming]{Inspecting Maude Variants with \Glints 
\thanks{This work has been partially supported by EU (FEDER) and Spanish MINECO grant TIN 2015-69175-C4-1-R and by Generalitat Valenciana PROMETEO-II/2015/013. A. Cuenca-Ortega is supported by SENESCYT, Ecuador (scholarship program 2013), J. Sapi\~na by FPI-UPV grant SP2013-0083, and S. Escobar by the Air Force Office of Scientific Research under award number FA9550-17-1-0286.}
}

\author{ M.~Alpuente\inst{1} \and A.~Cuenca-Ortega \inst{1,2} \and S.~Escobar\inst{1} \and J.~Sapi\~na\inst{1}}

\institute{
	DSIC-ELP, Universitat Polit\`ecnica de Val\`encia, Spain \\
	\email{\{alpuente,acuenca,sescobar,jsapina\}@dsic.upv.es}
	\and
	Universidad de Guayaquil, Ecuador \\
	\email{angel.cuencao@ug.edu.ec}
}

\maketitle

\begin{abstract}
This paper introduces \Glints, a graphical tool for exploring variant narrowing computations in Maude. The most recent version of Maude, version 2.7.1, provides quite sophisticated unification features, including order-sorted equational unification for convergent theories modulo axioms such as associativity, commutativity, and identity (ACU). This novel equational unification relies on built-in generation of the set of \emph{variants} of a term $t$, i.e., the canonical form of $t \sigma$ for a computed substitution $\sigma$. Variant generation relies on a novel narrowing strategy called \emph{folding variant narrowing} that opens up new applications in formal reasoning, theorem proving, testing, protocol analysis, and model checking, especially when the theory satisfies the \emph{finite variant property}, i.e., there is a finite number of most general variants for every term in the theory. However, variant narrowing computations can be extremely involved and are simply presented in text format by Maude, often being too heavy to be debugged or even understood. The \Glints~system provides support for (i) determining whether a given theory satisfies the finite variant property, (ii) thoroughly exploring variant narrowing computations, (iii) automatic checking of node {\em embedding} and {\em closedness} modulo axioms, and (iv) querying and inspecting selected parts of the variant trees. This paper is under consideration for acceptance in TPLP.
\end{abstract}
\pagestyle{plain}

\section{Introduction}
Narrowing is a symbolic execution mechanism that generalizes term rewriting by allowing free variables in terms (as in logic programming) and handles them by using unification (instead of pattern matching) to non-deterministically reduce these terms. Originally introduced in the context of theorem proving, narrowing is complete in the sense of logic programming (computation of answers) and functional programming (computation of irreducible forms) so that efficient versions of narrowing have been adopted as the operational principle of so-called multi-paradigm (functional logic) programming languages (see, e.g., \cite{Hanus13}). In the last few years, there has been a resurgence of narrowing in many application areas such as equational unification, state space exploration, protocol analysis, termination analysis, theorem proving, deductive verification, model transformation, testing, constraint solving, and model checking. To a large extent, the growing interest in narrowing is motivated by the recent takeoff of symbolic execution applications and the availability of efficient narrowing implementations.

Maude is a language and a system that efficiently implements Rewriting Logic \cite{Meseguer92}, which is a logic of change that seamlessly unifies a wide variety of models of concurrency. Thanks to its logical basis, Maude provides a precise mathematical model, which allows it to be used as a declarative language and as a formal verification system. The most recent version of Maude, version 2.7.1~\cite{maude-manual}, provides quite sophisticated narrowing-based features, including order-sorted equational unification for convergent theories modulo a set of commonly occurring axioms such as associativity, commutativity, and identity (ACU). This novel equational unification relies on built-in generation of the set of variants of a term $t$~\cite{DEEM+16}. A \emph{variant} \cite{CD05} of a term $t$ in the theory ${\cal E}$ is the canonical (i.e.,\ irreducible in ${\cal E}$) form of $t \sigma$ for a given substitution $\sigma$; in symbols, $(\sigma, t\sigma\!\!\downarrow_{{\cal E}}$). Variants are computed in Maude by using the \emph{folding variant narrowing strategy}~\cite{ESM12}, which adopts from \emph{tabled logic programming} \cite{CW96} the idea of memoizing calls encountered in a query evaluation (along with their answers) in a set of tables so that, if the call is re-encountered, the information from the table is reused instead of running the call again. This is useful in two ways: it prevents looping, which may ensure termination under suitable conditions, and it filters out redundant derivations to a reachable expression leading to better performance. When a convergent theory satisfies the \emph{finite variant property} (i.e., there is a finite number of most general variants for every term in the theory), folding variant narrowing computes a minimal and complete set of most general variants in a finite amount of time. Many theories of interest have the FVP, including theories that give algebraic axiomatizations of cryptographic functions used in communication protocols, where FVP is omni-present.

Maude's variant generation mechanism was originally designed as an aid for order-sorted equational unification modulo axioms and related problems. It delivers the set of most general variants of the given theory, but it does not allow the user to control the process in any way nor does it provide the user with thorough information about the variant computation process. Unfortunately, variant computations delivered by Maude using the folding variant narrowing strategy can be extremely involved and are simply presented in text format, often being too heavy to be debugged or even understood. 

Recently, the definition and inspection of equational theories for which the variants are generated has become an interesting application on its own, which requires enhanced support for exploring the variant narrowing computations. For example, \cite{YEMMN14} considers twenty equational theories for protocol analysis in the protocol analyzer Maude-NPA. These equational theories represent under- and over-approximations of the theory of homomorphic encryption with different variant generation behaviors (see \cite{YEMMN14} for details). As another example, \cite{Meseguer15} considers distinct axiomatizations of several equational theories of interest for boolean satisfiability. Given the huge intricacy of variant computations, in both cases the development of all these equational theories was painful when considering the time and effort required to analyze the different variant-based properties for the considered versions of the theories. Often, even an ordinary developer who uses (variant) narrowing as a functional--logic program execution mechanism needs deeper support than currently provided by Maude.

This paper describes an inspecting tool for variant computations in Maude called \Glints~\linebreak(\textit{GraphicaL Interactive Narrowing Tree Searcher)} and its implementation. \Glints\ does not only visualize the variants generated by Maude; it goes beyond that by showing internal narrowing computations in full detail, including partially computed substitutions, $Ax$-matching and equational normalization steps that are concealed within Maude's variant narrowing and equational rewriting algorithms. Exploration and visualization in \Glints~can be either automatic or interactive, which allows following promising paths in the narrowing tree without exploring irrelevant parts of it. This supports the design of efficient heuristics for some applications. Also, the displayed view can be abstracted when its size requires it, to avoid cluttering the display with unneeded details. Important insights regarding the programs/theories can be gained from controlling the narrowing space exploration. Does the theory have a finite number of variants? How many variants are there? What do these variants look like and how do they compare to each other modulo axioms? (For instance, is one of the nodes {\em embedded} or structurally subsumed by one of its ancestors? Is the node {\em closed} or an equational instance of the tree root or input expression?) What is the meta-level representation of a narrowing computation trace? Moreover, it can also help uncover correctness bugs or even unexpected low performance (by showing which patterns have been executed more often or dominate the execution), which might otherwise be very difficult to identify. As far as we know, this is the first graphical tool in the literature for visual inspection of variant narrowing computations modulo axioms.

After introducing the basic ideas of Maude's narrowing-based variant generation in Section \ref{sec:variant-nwing}, we introduce a leading example for describing \Glints~equational reasoning capabilities based on variant narrowing in Section \ref{sec:example}. We explain the core functionality of \Glints~and extra inspection features in Section \ref{sec:glimpse}. We provide a description of the tool implementation together with some experiments that assess its performance in Section \ref{sec:imple}. Finally, some related work and further applications are briefly discussed in Section \ref{sec:related}.

\Glints\ is publicly available at \url{http://safe-tools.dsic.upv.es/glints}. 

\section{Narrowing-based variant generation in Maude}\label{sec:variant-nwing}
Unification is a deductive mechanism that is used in many automated deduction tools and is essential for programming languages with logical capabilities. Although Maude inherited many features from predecessor languages, like OBJ and Eqlog \cite{Meseguer06}, for the sake of high-performance, the narrowing-based, logic programming capabilities of the equational logic language Eqlog were left behind since the first public release of Maude in 1999. Order-sorted unification and narrowing modulo axioms first became available in 2009 as a part of Maude 2.4, while variant generation, variant-based unification, and folding variant narrowing have only been implemented twenty years later as built-in, highly efficient features in Maude 2.7.1. 

Let us illustrate the notion of narrowing in Maude by considering the following simple Maude\footnote{Maude syntax is almost self-explanatory, using explicit keywords such as \texttt{fmod}, \texttt{sort}, and \texttt{op} to respectively introduce a module, a sort (or type), and an operator, together with its {\em domain} $\rightarrow$ {\em range} typing declaration that appears after the `:' symbol (e.g., {\tt op s : Nat -> Nat}). The sort of a variable can be given explicitly in any expression or within the variable declaration section. In addition, from Maude 2.7 and later, only equations with the attribute \texttt{variant} are used by the folding variant narrowing strategy, while all of the others are only used for rewriting.} functional module (with no axioms $Ax$) for addition \texttt{NAT-VARIANT}. 
 
{\footnotesize
\begin{verbatim} 
  fmod NAT-VARIANT is
    sort Nat .
    op 0 : -> Nat .
    op s : Nat -> Nat .
    op _+_ : Nat Nat -> Nat .
    vars X Y Z W : Nat .
    eq [1] : 0 + Y = Y [variant] .
    eq [2] : s(X) + Y = s(X + Y) [variant] .
  endfm
\end{verbatim} 
}

The \emph{red}ucible \emph{ex}pression, or simply {\em redex}, \texttt{s(0)+0} can be simplified into the result \texttt{s(0)} in two rewriting steps as follows: \texttt{s(0)+0} $\!\rightarrow_{\texttt{[2]}}\!$ \texttt{s(0+0)} $\!\rightarrow_{\texttt{[1]}}\!$ \texttt{s(0)}. Similarly, the non-ground term \texttt{s(0)+W} can be simplified into the result \texttt{s(W)} in two rewriting steps as follows: \texttt{s(0)+W} $\!\rightarrow_{\texttt{[2]}}\!$ \texttt{s(0+W)} $\!\rightarrow_{\texttt{[1]}}\!$ \texttt{s(W)}. However, even though the term \texttt{Z+0} cannot be rewritten as it does not match the left-hand side (lhs) of any equation, it can be {\em narrowed} into \texttt{s(0)}, with {\em computed answer substitution} $\theta = \{\texttt{Z}\mapsto\texttt{s(0)}\}$, in two narrowing steps as follows: \texttt{Z+0} ${\stackrel{\sigma_0}{\leadsto}}_{\texttt{[2]}}$ \texttt{s(X+0)} ${\stackrel{\sigma_1}{\leadsto}}_{\texttt{[1]}}$ \texttt{s(0)}, where ${\sigma_0}=\{\texttt{Z}\mapsto\texttt{s(X)}, \texttt{Y}\mapsto\texttt{0}\}$ is the most general unifier (mgu) of \texttt{Z+0} and the lhs \texttt{s(X)+Y} of the applied equation \texttt{[2]}, and similarly ${\sigma_1}=\{\texttt{X}\mapsto\texttt{0},\texttt{Y'}\mapsto\texttt{0}\}$ is the mgu of \texttt{X+0} and the (renamed-apart) lhs \texttt{0+Y'} of equation [1], and $\theta = (\sigma_{0}\sigma_{1})_{\res{\{Z\}}}$.

For an equational theory $\mathcal{E}=(\Sigma, E \cup Ax)$ to be executable, its equations $E$ must be {\em convergent} (i.e., confluent, terminating, and coherent modulo the given axioms $Ax$) \cite{DM12}. This ensures: 1) that every input expression $t$ has one (and only one) {\em canonical} form $t\!\!\downarrow_{{\cal E}}$\/; and 2) that the Maude interpreter can implement conditional rewriting $\rightarrow_{E \cup Ax}$ as a much simpler relation $\rightarrow_{E, Ax}$ (rewriting with $E$ modulo $Ax$) that uses the equations of $E$ (oriented from left to right) as the only simplification rules, while the equations in $Ax$ are just encapsulated within a powerful algorithm of pattern matching modulo $Ax$ that is used at each rewrite step with $E$.

Given ${\cal E} = (\Sigma, E \cup Ax)$, the $(E, Ax)$-\emph{variants} of a term $t$ are the set of all pairs $(\sigma, t\sigma\!\!\downarrow_{\cal E})$, each one of which consists of a substitution $\sigma$ and the $(E, Ax)$-canonical form of $t\sigma$ \cite{CD05,ESM12}. Intuitively, the variants of $t$ are the ``irreducible patterns'' to which $t$ can be symbolically evaluated by applying the (implicitly oriented) equations of $E$ modulo $Ax$. For instance, there is an infinite number of variants for the term \texttt{(0 + Y:Nat)} in the theory {\tt NAT-VARIANT}, e.g., $(id,\texttt{Y:Nat})$, $(\{\texttt{Y:Nat}\mapsto\texttt{0}\},\texttt{0})$, $(\{\texttt{Y:Nat}\mapsto\texttt{s(Z:Nat)}\},\texttt{s(Z:Nat)})$, $(\{\texttt{Y:Nat}\mapsto\texttt{s(0)}\},\texttt{s(0)})$, $\ldots$.

A preorder relation of generalization between variants provides a notion of {\em most general variant} and also a notion of completeness of a set of variants. For the term \texttt{0 + Y:Nat}, the most general variant is $(id,\texttt{Y:Nat})$ since any other variant can be obtained by equational instantiation. 

For example, consider the following theory that declares the two Boolean constants \texttt{true} and \texttt{false}. The key thing to note are the special attributes \texttt{assoc} and \texttt{comm}, meaning that the infix operators ``\texttt{and}'' and ``\texttt{or}'' obey associativity and commutativity axioms:

{\footnotesize
\begin{verbatim} 
  fmod BOOL is 
    sort Bool .
    ops true false : -> Bool .
    ops _and_ _or_ : Bool Bool -> Bool [assoc comm] .
    vars X Y : Bool .
    eq X and true = X [variant] .
    eq X and false = false [variant] .
    eq X or true = true [variant] .
    eq X or false = X [variant] .
  endfm
\end{verbatim} 
}

\noindent There are five most general variants modulo AC for ``\texttt{X and Y}'', which are: $\{(id, \texttt{X and Y}), \linebreak(\{\texttt{X} \mapsto \texttt{true}\}, \texttt{Y}), (\{\texttt{Y}\mapsto \texttt{true}\}, \texttt{X}), (\{\texttt{X} \mapsto \texttt{false}\}, \texttt{false}), (\{\texttt{Y} \mapsto \texttt{false}\}, \texttt{false})\}$. Similarly, there are five most general variants for ``\texttt{X or Y}''.

An equational theory has the \emph{finite variant property} (FVP) (or it is called a \emph{finite variant theory}) iff there is a finite and complete set of most general variants for each term. The theory BOOL is FVP, whereas the theory \texttt{NAT-VARIANT} does not have the finite variant property since there is an infinite number of variants in \texttt{NAT-VARIANT} for the term \texttt{X:Nat + 0}. It is generally undecidable whether an equational theory has the FVP ~\cite{BGLN13}; a semi-decision procedure is given in \cite{Meseguer15} that works well in practice and another technique based on the dependency pair framework is given in \cite{ESM12}. The procedure in \cite{Meseguer15} works by computing the variants of all flat terms $f(X_1,\ldots,X_n)$ for any $n$-ary operator $f$ in the theory and pairwise-distinct variables $X_1,\ldots,X_n$ (of the corresponding sort); the theory does have the FVP iff there is a finite number of most general variants for every such term \cite{Meseguer15}. 

At the practical level, variants are generated by using an efficient narrowing strategy called the \emph{(folding) variant narrowing strategy}, which was proved to be complete for variant generation in \cite{ESM12} and terminates for all inputs provided that the theory has the FVP. Variant narrowing derivations correspond to narrowing sequences $t_0~ {\stackrel{\sigma_0}{\leadsto}}_{e_0,Ax} ~t_1 ~{\stackrel{\sigma_1}{\leadsto}}_{e_1,Ax} ... {\stackrel{\sigma_{n-1}}{\leadsto}}_{e_{n-1},Ax} ~ t_n,$ where $t {\stackrel{\sigma}{\leadsto}}_{e,Ax} ~t'$ denotes a transition (modulo axioms in $Ax$) from term $t$ to $t'$ via the {\em variant equation} $e$ (i.e., an equation $e$ that is enabled to be used for narrowing thanks to the attribute \texttt{variant}) using the {\em equational unifier} $\sigma$. Assuming that the initial term $t$ is normalized, each single transition $t {\stackrel{\sigma}{\leadsto}}_{e,Ax} ~t'$ (or variant narrowing step) is followed by the simplification of the term into its normal form by using all equations in the theory, which may include not only the variant equations in the theory but also (non-variant) equations (e.g., built-in equations in Maude). The composition $\sigma_0 \sigma_1 \sigma_{n-1}$ of all the unifiers along a narrowing sequence leading to $t_n$ (restricted to the variables of $t_0$) is the {\em computed variant substitution} of this sequence. The {\em folding} refinement of variant narrowing that is implemented in Maude essentially consists in ``folding'', by subsumption modulo $Ax$, the narrowing tree for (${E},Ax$), which can in practice result in a finite narrowing graph that symbolically summarizes the, in general infinite, ($E,Ax$)-narrowing tree.

Maude provides the following command for variant generation: 

\noindent
{\small
\begin{alltt}
  get variants [ n ] in \nt{ModId} : \nt{Term} .
\end{alltt}
}

\noindent where $n$ is an optional argument that indicates the number of variants requested and \nt{ModId} is the module where the command is run. There is also a meta-level command for variant generation, see \cite{maude-manual}.

For example, consider the following equational theory \cite{maude-manual} for the {\em exclusive-or} symbol \texttt{\_*\_} (i.e., an exclusive union operator \texttt{\_*\_} for sets of natural numbers, {\tt NatSet}, such that $X1 * X2$ is the set of natural numbers appearing in $X1$ or $X2$, but not both), where \texttt{mt} is the (empty set) identity element. Note that the notation \texttt{[NatSet]} denotes the {\em kind} of sort \texttt{NatSet} that, in addition to normal data of sort \texttt{NatSet}, can also contain ``error expressions''.

{\footnotesize
\begin{verbatim}
  fmod EXCLUSIVE-OR is 
    sorts Nat NatSet .  subsort Nat < NatSet .
    op 0 : -> Nat .
    op s : Nat -> Nat .
    op mt : -> NatSet .              
    op _*_ : NatSet NatSet -> NatSet [assoc comm] .
    vars X Z : [NatSet] .
    eq [idem] :     X * X = mt    [variant] .
    eq [idem-Coh] : X * X * Z = Z [variant] .
    eq [id] :       X * mt = X    [variant] .
  endfm
\end{verbatim}
}

\noindent We can check that the above theory has the finite variant property by asking Maude to generate all variants for the exclusive-or symbol $*$ in the \texttt{EXCLUSIVE-OR} module, which delivers 7 variants:

{\footnotesize
\begin{verbatim}
  Maude> get variants in EXCLUSIVE-OR : X * Y .
  Variant 1                            Variant 2                   Variant 7
  [NatSet]: #1:[NatSet] * #2:[NatSet]  NatSet: mt    .........     [NatSet]: %1:[NatSet]
  X --> #1:[NatSet]                    X --> #1:[NatSet]           X --> %1:[NatSet]
  Y --> #2:[NatSet]                    Y --> #1:[NatSet]           Y --> mt
\end{verbatim}
}

\noindent Observe that Maude can introduce fresh variables of two classes: (\verb|#|\texttt{$n{:}Sort$}) or (\texttt{$\%n{:}Sort$}). This is because it distinguishes between variables that are generated by the built-in unification modulo axioms (\verb|#|\texttt{$n{:}Sort$}) and variables that are generated by variant-based unification or variant generation (\texttt{$\%n{:}Sort$}) \cite{maude-manual}. Also note that the canonical form for any other instance of the term \texttt{X * Y} is subsumed modulo the axioms by one of the seven computed variants. For instance, when the substitution $\{\texttt{X}\mapsto \texttt{0 * s(0)},\texttt{Y}\mapsto \texttt{0 * s(0)}\}$ is applied to \texttt{X * Y}, the canonical form is just \texttt{mt}, but this is an instance of Variant $2$. This is because the application (modulo associativity and commutativity of \texttt{*}) of equation \texttt{[idem]} causes \texttt{0} and \texttt{s(0)} to be cancelled. Other examples of variant generation can be found in the Maude manual~\cite{maude-manual}. Maude can also be asked to return the sequence of most general variants \emph{incrementally}, which can be useful when a theory does not have the finite variant property. For instance, the term \texttt{X:Nat + s(0)} has an infinite number of most general variants in the theory of the module \texttt{NAT-VARIANT}. In such a case, Maude can either output the infinite sequence of variants to the screen (and the user can stop the process whenever she wants), or be given a bound $n$ so that it generates only a maximum of $n$ variants.

{\footnotesize
\begin{verbatim}
  Maude> get variants [10] in NAT-VARIANT : X + s(0) .
  Variant 1                                              Variant 10
  Nat: #1:Nat + s(0)    ............................     Nat: s(s(s(s(s(0)))))
  X --> #1:Nat                                           X --> s(s(s(s(0))))
\end{verbatim}
}

\noindent
In the case when the bound $n$ is reached, the user can incrementally increase the bound so that the FVP is proved whenever the number of computed variants is smaller than the given bound. Unfortunately, the variant generation process can be infinitely repeated if the FVP does not hold.

In the following section, we show how proving that a theory has the FVP is much easier and fruitful by using \Glints. Actually, we might even know that the FVP is not fulfilled and yet be interested in exploring the variant narrowing computation space of a number of terms in order to gain insights on how to modify the theory so that the FVP holds.

\section{Folding variant narrowing trees in \Glints: a running example}\label{sec:example}
Let us consider again the equational specification for the exclusive-or theory above. This theory has the FVP since only seven most general variants exist for the symbol \verb!_*_!. However, one might be interested to grasp why this specification fulfills the FVP, whereas slightly modified specifications of the exclusive-or theory are known to fail. 

For example, assume that we test the FVP after replacing the variable declaration \linebreak\texttt{X:[NatSet]} of the original specification with \texttt{X:Nat}:

{\footnotesize
\begin{verbatim}
  fmod EXCLUSIVE-OR-NOFVP is 
    sorts Nat NatSet .  subsort Nat < NatSet .
    op 0 : -> Nat .
    op s : Nat -> Nat .
    op mt : -> NatSet .
    op _*_ : NatSet NatSet -> NatSet [assoc comm] .
    var X : Nat . var Z : [NatSet] .
    eq [idem] :     X * X = mt    [variant] .
    eq [idem-Coh] : X * X * Z = Z [variant] .
    eq [id] :       X * mt = X    [variant] .
  endfm
\end{verbatim}
}

\begin{figure}[t]
\centering
\includegraphics[width=\linewidth]{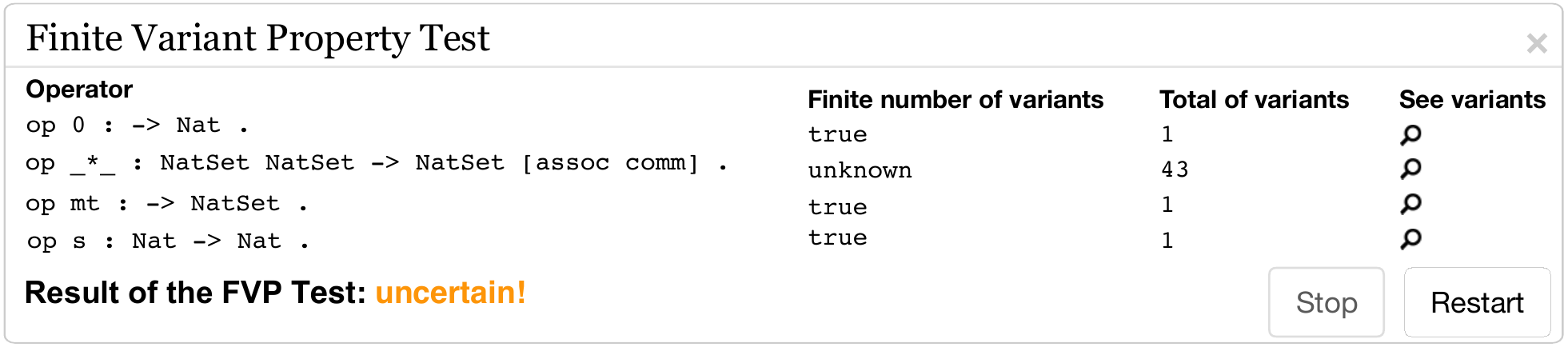}
\caption{The FVP test for the modified non-FVP exclusive-or theory.}\label{fig:xor-noFVP-testFVP}
\end{figure}

\begin{figure}[t]
\centering
\includegraphics[width=\linewidth]{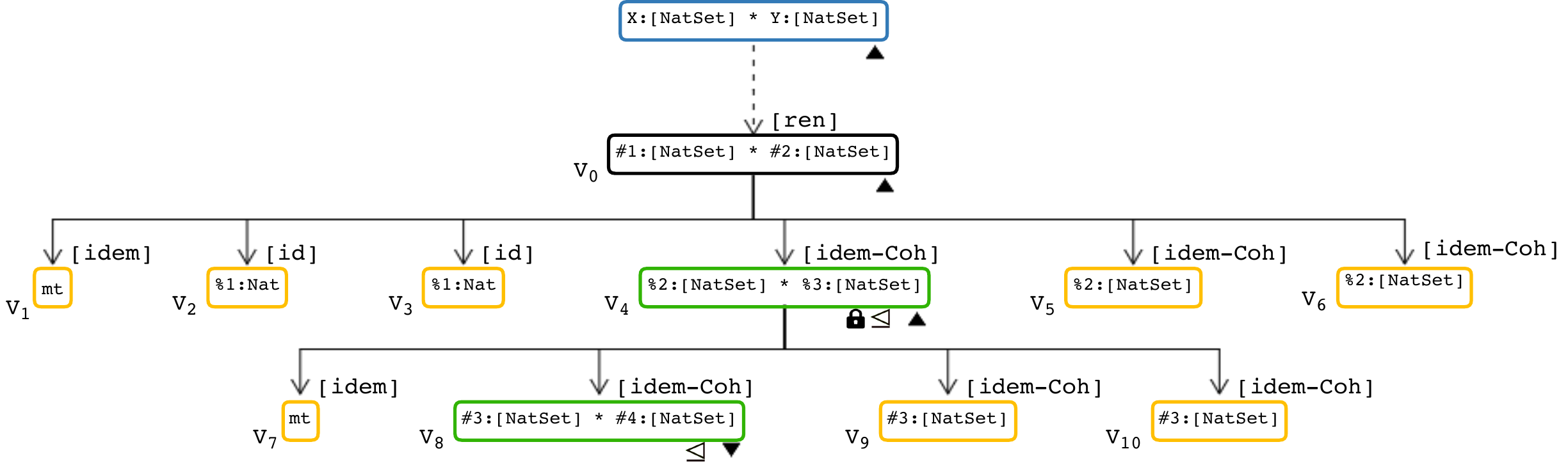}
\caption{Inspecting variant computations of the modified non-FVP exclusive-or theory.}\label{fig:xor-noFVP-embedding-and-closed}
\end{figure}

\noindent The variant generation process in Figure~\ref{fig:xor-noFVP-testFVP} is stopped after computing $43$ variants for symbol \verb!_*_! due to timeout, hence the result of the FVP test is {\em uncertain} yet this theory is known not to satisfy the FVP. One could investigate why this simple modification destroys FVP by inspecting the folding variant narrowing tree for the expression {\small\verb!X:[NatSet] * Y:[NatSet]!} shown in Figure~\ref{fig:xor-noFVP-embedding-and-closed}.

\Glints~can generate the folding variant narrowing tree of a given term in three ways: (i)~stepwisely, by (manually) selecting a down triangle symbol $\blacktriangledown$ that is shown below each narrowable node of the tree (see Figure~\ref{fig:xor-noFVP-embedding-and-closed}); (ii) automatically until a fixed depth bound is reached, or (iii) automatically by using the more sophisticated control mechanism called {\em (equational) homeomorphic embedding} that is commonly used to ensure termination of unfolding-based program transformation and other symbolic methods \cite{Leuschel02,ACEM17}. Informally, a term $t'$ embeds\footnote{The order-sorted extension of homeomorphic embedding modulo equational axioms, such as associativity, commutativity, and identity that we use for Maude can be found in \cite{ACEM17}.} another term $t$, in symbols $t \trianglelefteq t'$, if $t$ (or a term that is equal to $t$ modulo $Ax$) can be obtained from $t'$ by deleting some symbols of $t'$; e.g., {\tt s(s(X+Y)$\ast$(s(X)+Y))} embeds {\tt s(Y$\ast$(X+Y))}, assuming commutativity of the \verb!_*_! symbol. Nodes in the folding variant narrowing tree that embed a previous node in the same branch of the tree are highlighted in green and are decorated with symbol $\trianglelefteq$ below the node, as shown in Figure ~\ref{fig:xor-noFVP-embedding-and-closed} (by clicking on the symbol, its closest embedded ancestor gets also highlighted).

In Figure~\ref{fig:xor-noFVP-embedding-and-closed}, note that we have interactively produced variants up to $V_{10}$ and could continue generating variants indefinitely whereas the folding variant narrowing tree for the original \linebreak\texttt{EXCLUSIVE-OR} theory stops at node $V_6$. Also note that some potential narrowing steps stemming from the nodes of Figure \ref{fig:xor-noFVP-embedding-and-closed} are not produced by the folding variant narrowing strategy as it avoids expanding nodes that are subsumed by previous ones. For instance, for node $V_4$, folding variant narrowing does not compute any children nodes equivalent to children $V_2$ and $V_3$ of node $V_0$. However, the theory \texttt{EXCLUSIVE-OR-NOFVP} does not have the FVP because nodes $V_7,V_8,V_9,V_{10}$ are not subsumed by their counterpart nodes $V_1,V_4,V_5,V_6$, respectively, whereas they are subsumed for the theory \texttt{EXCLUSIVE-OR}, yielding the seven variants $V_0,\ldots,V_6$.

\begin{figure}[t]
\centering
\includegraphics[width=\linewidth]{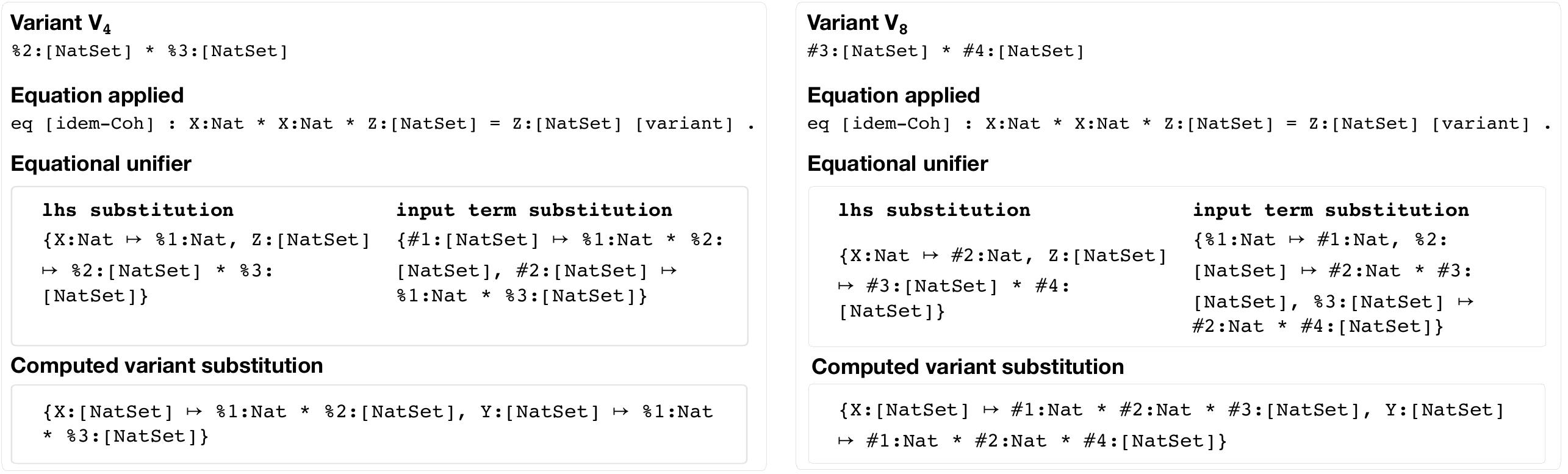}
\caption{Comparison of nodes $V_4$ and $V_8$.}\label{fig:Variant-comparison}
\end{figure}

The fact that \Glints~automatically detects that node $V_0$ in Figure \ref{fig:xor-noFVP-embedding-and-closed} is trivially \emph{embedded} into node $V_4$, and that $V_4$ is \emph{embedded} into node $V_8$ (actually they are all equal modulo variable renaming), warns about potentially infinite narrowing computations stemming from $V_0$ (it is said that $\trianglelefteq$ {\em whistles} \cite{Leuschel02}). However, note that node $V_8$ is not a variant of $V_4$ (nor $V_{0}$). By comparing nodes $V_4$ and $V_8$ (enabled by pressing \textit{Compare nodes} in the top-right menu), we obtain the information of Figure~\ref{fig:Variant-comparison}, which reveals that, even though $V_4$ and $V_{8}$ are equal modulo renaming, the computed variant substitutions are different. 

After considering a negative example where \Glints~could help you to understand when and why the finite variant property of a theory can be lost after some changes, let us now analyze a positive example where an equational specification can satisfy the FVP after some changes.

\begin{figure}[t]
\centering
\includegraphics[width=.8\linewidth]{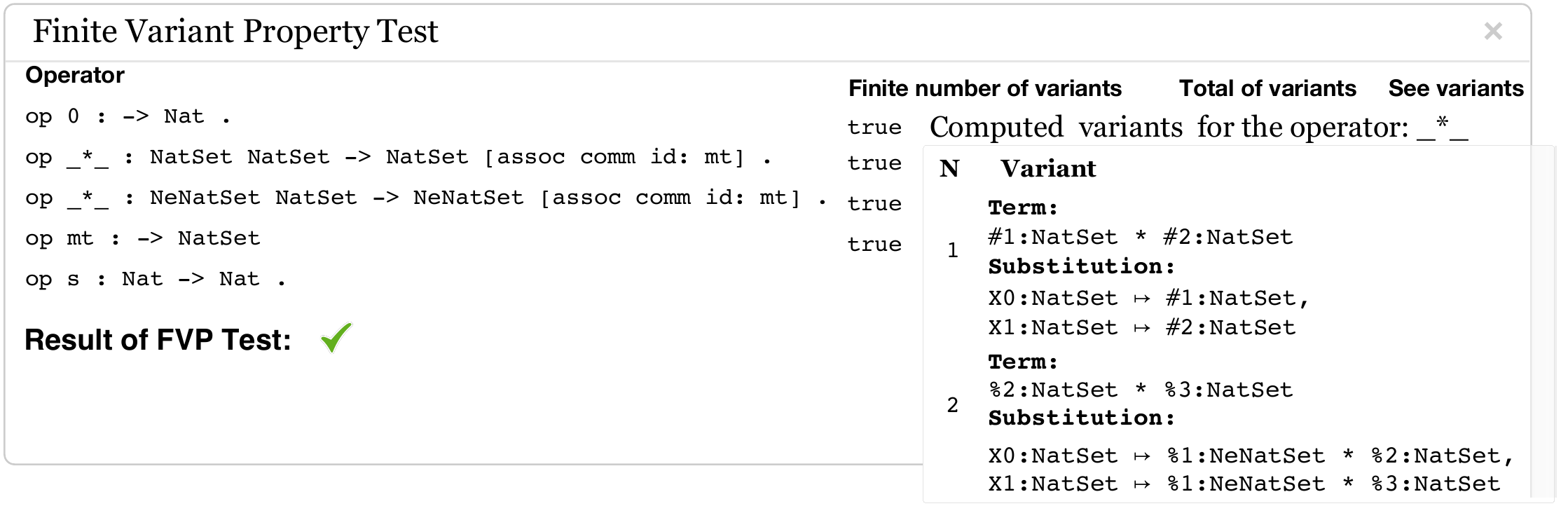}
\caption{FVP test for the newly modified exclusive-or theory with true verdict and variants.}
\label{fig:xor-ACU-testFVP}

\end{figure}

If we make the original specification and make the \verb!_*_! symbol be associative, commutative, and with identity empty set element \texttt{mt}, then the theory does have FVP. This is shown in Figure~\ref{fig:xor-ACU-testFVP}; the list of computed variants for the operator \verb!_*_! symbol is also shown, which has been retrieved by simply clicking on the corresponding \includegraphics[width=1.5ex]{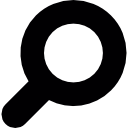} symbol of the right column. 
 
{\footnotesize
\begin{verbatim}
  fmod EXCLUSIVE-OR-ACU is 
    sorts Nat NeNatSet NatSet .  subsort Nat < NeNatSet < NatSet .
    op 0 : -> Nat .    
    op s : Nat -> Nat .    
    op mt : -> NatSet .
    op _*_ : NatSet NatSet -> NatSet [assoc comm id: mt] .
    op _*_ : NeNatSet NatSet -> NeNatSet [assoc comm id: mt] .
    var X : NeNatSet .     var Z : [NatSet] .
    eq [idem-Coh] : X * X * Z = Z [variant] .
  endfm
\end{verbatim}
}

\noindent Note that this new specification relies on a subsort relation between sets of natural numbers (sort {\tt NatSet}) and non-empty sets of natural numbers (sort {\tt NeNatSet}), and it is simpler than the previous one because only one equation is needed. If variable \texttt{X} were given the sort \texttt{Nat} instead of \texttt{NeNatSet}, the mutated theory would not satisfy the FVP. 

Folding variant narrowing trees can also be checked in \Glints~for the {\em (equational) closedness} property, which naturally extends to order-sorted equational theories (being executed by folding variant narrowing) the standard notion of closedness\footnote{This notion was generalized to the narrowing-driven partial evaluation of functional-logic programs that are modeled as (unsorted) term rewriting systems in \cite{AFV98,AFMV97}.} of program calls that is used in the partial deduction (PD) of logic programs, meaning that the call is an instance of one of the specialized expressions. \Glints\ implements the equational closedness check for the nodes of the deployed folding variant narrowing tree w.r.t.\ the root of the tree; this transfers to our setting the idea of {\em regularity} of a symbolic computation (in the terminology of \cite{AFV98b,PP96}). Informally, a node in the tree is equationally closed (w.r.t.\ the tree root) if each narrowing redex in the term is an {\em equational instance} of the root node of the tree. For instance, for a tree with root {\tt (X*Y)} and with one leaf node {\tt (0*X*Z)}, and assuming associativity of \verb!_*_!, there are three redexes (namely, {\tt (0*X)}, {\tt (X*Z)}, and {\tt (0*X*Z)}) and the leaf is closed. Note that neither closedness implies embedding nor the opposite: {\tt (0*0)} is closed w.r.t. {\tt (X*Y)} yet it does not embed it, and {\tt (0*X*Y)} embeds {\tt (X*X)} yet it is not closed w.r.t.\ it.

It is interesting to note that the notion of variant is closely related to the (functional-logic) notion of {\em resultant\/} that is used in unfolding-based symbolic transformation techniques that rely on (some form of) narrowing, such as the narrowing-driven partial evaluator for TRSs of \cite{AFV98} and the partial evaluator {\sf Victoria} for Maude equational theories of \cite{ACEM17}, which is based on folding variant narrowing: given a narrowing tree for the term $t$ in the equational theory ${\cal E}$, for each ({\em root--to--leaf\/}) narrowing derivation $t \leadsto^{*}_{\sigma} s$ in the tree, specialized (oriented) equations $t\sigma = s$ (also called {\em resultants}) can be extracted from the tree by piecing together the last term $s$ of the narrowing derivation with the corresponding instance $t\sigma$ of the initial term $t$. Similarly to PD, in the partial evaluator of \cite{ACEM17}, {\em equational closedness} is the key property to ensure that, given a set $Q$ of input expressions, the set resultants that can be extracted from a set of folding variant narrowing trees built in ${\cal E}$ for the terms of $Q$ (each one as explained above) form a complete description that {\em correctly specializes} the original theory ${\cal E}$ to the considered set $Q$. In other words, all calls that may occur at run-time when any instance (modulo $Ax$) of a term of $Q$ is executed in the specialized theory $\cal S$ are {\em covered} by $\cal S$ (i.e.,\ folding variant narrowing computes the same solutions in $\cal S$ as in the original theory $\cal E$).
 
Similarly to the equational embedding test, the equational axioms and the order-sortedness information are both considered in the equational closedness test that is implemented in \Glints. The tool checks this property automatically at any node in which the homeomorphic embedding whistles, and also when a node is interactively selected. It is signaled by an extra symbol \includegraphics[width=1.5ex]{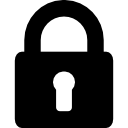} below the node (except for unnarrowable leaves, which are always trivially closed and are simply highlighted in orange). In Figure~\ref{fig:xor-noFVP-embedding-and-closed} all the nodes are equationally closed; actually, they are either unnarrowable or a syntactic instance of the tree root.

\vspace{-0.5cm}

\section{\Glints~at a glimpse}\label{sec:glimpse}
In this section, the main features of the graphical explorer \Glints~are outlined. Once a Maude module (or sequence of modules) has been input, the initial \Glints~panel allows: 1)~the folding variant narrowing space to be explored for a given term; and 2)~the finite variant property to be checked (as explained in Section \ref{sec:example}).

Running the graphical explorer and executing the corresponding textual narrowing commands of Maude is essentially the same regarding the processes that are conducted in the background (i.e.,\ to some extent, the narrowing tree panel can be interpreted as the visual correspondent of the \verb^show-search-graph^ command from the textual narrower). However, there is a dramatic difference in the tool output and in the thoroughness of the reasoning support provided by \Glints. 

\vspace{-0,35cm}

\subsection{Interactive tree unfolding and querying}

Given an input term, the graphical narrowing tree panel initially contains two nodes: the input node and its normalized version w.r.t.\ the theory. Additions to the graph will be dictated by the user's exploration actions, which can be as follows.

\vspace{-.25cm}

\paragraph{\textbf{\textit{Interactive exploration.}}}
\Glints~offers a graphical representation of the variant narrowing trees, including at each step (i) the narrowing redex, (ii) the applied variant equation, (iii) the equational unifier, and (iv) the computed variant substitution. \Glints~allows the narrowing tree to be easily navigated while providing thorough information regarding every node and edge in the tree. This is particular useful for a rich language such as Maude that supports sorts, subsorts and overloading, and equational rewriting modulo axioms such as ACU, where intuition is easily lost.

Each variant node is identified with a tag $V_n$, where $n$ is the variant number assigned by Maude. When a node is selected (by a simple click), it is shaded in yellow so that the user can be constantly aware of the current selection. Node selection is useful for centering the node inside the tree layout and is also used for checking the equational closedness property. Fully detailed information about each variant can be displayed by double-clicking on the corresponding node. Multiple variant information windows can be opened without updating the current tree. 

As is common in visualization tools, the search trees can be scaled and subtrees can be hidden. This is done by pressing the $\blacktriangle$ symbol that is displayed below each node. By doing so, the entire (sub-)tree (except for its root) is removed from the displayed view of the tree. Taking into account that the size of the tree can become considerably large, zooming capabilities are also enabled.

\vspace{-.25cm}

\begin{figure}[t]
	\centering
	\includegraphics[width=.65\linewidth]{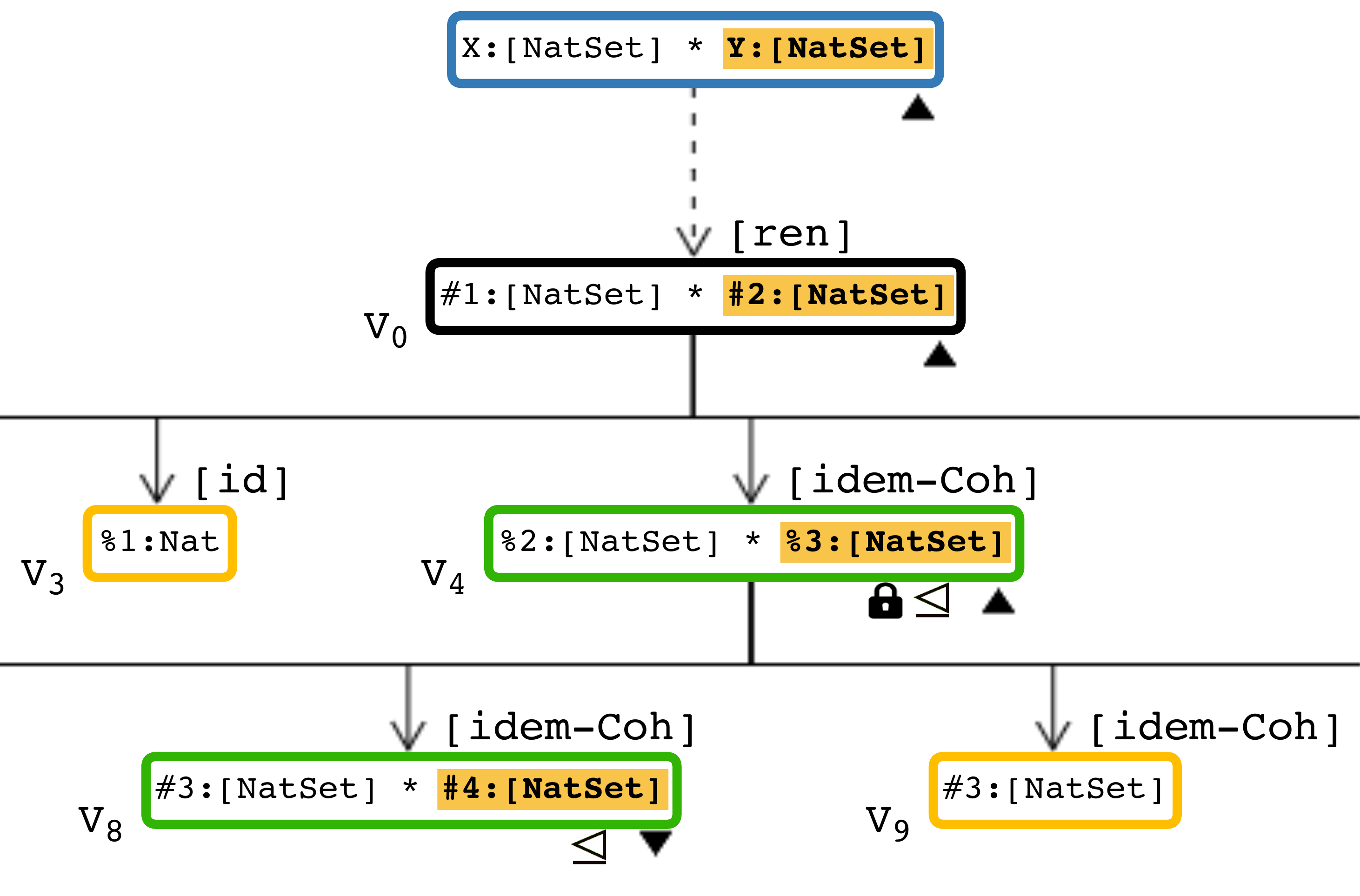}
	\caption{Result of the query ``\texttt{\_ * ?}'' for the VNT of the non-FVP exclusive-or theory.}\label{fig:search}
	\vspace{-0.25cm}
\end{figure}

\paragraph{\textbf{\textit{Tree querying.}}}
A querying box is displayed at the bottom of the narrowing tree panel that allows information of interest to be easily searched in huge narrowing trees by undertaking a query that specifies a template for the search. A query is a filtering pattern with wildcards that define irrelevant symbols by means of the underscore character ({\tt \underline{\phantom{m}}}) and define relevant symbols by means of the question mark character ({\tt ?}). For instance, asking the query ``{\tt \_ * ?}'' in the tree of Figure~\ref{fig:xor-noFVP-embedding-and-closed} highlights expressions {\small\verb!#2:[NatSet]!}, {\small\verb!%3:[NatSet]!}, and {\small\verb!#4:[NatSet]!} in nodes $V_0$, $V_4$, and $V_8$, respectively, as shown in Figure \ref{fig:search}.

\vspace{-0,25cm}

\subsection{Automated tree unfolding, enriched views and exporting}

By using \Glints, variant generation can be easily automated in multiple ways. Specifically, the user can ask the searcher to do one of the following: (i) deliver the first $n$ variants of the considered initial term, (ii) compute the entire narrowing tree up to a given depth, or (iii) compute the entire narrowing tree until the embedding whistles along all branches. In all cases, exploration of the tree stops whenever the corresponding termination criterion is met, namely (i) no more variants exist, (ii) bound is reached, (iii) embedding whistles, or (iv) timeout is surpassed.

By clicking the $\equiv$ symbol that appears in the right corner of the window, a command menu is displayed that automates these capabilities by means of the following accessible buttons.

\vspace{-.15cm}

\paragraph{\textbf{\textit{Depth-$k$ (resp. $N$-variants) expansion.}}} It unfolds the tree automatically down to its depth-$k$ frontier (resp. until the $n$-th variant has been computed). An input box allows one to fix the desired {\em upper} bound in the depth of the tree or in the number of solutions. 

\vspace{-.15cm}

\paragraph{\textbf{\textit{Embedding-based expansion.}}} It automatically unfolds the variant narrowing tree by relying on equational homeomorphic embedding to ensuring finiteness. Roughly speaking, whenever a new node $t_{n+1}$ is to be added to a branch, \Glints\ checks whether $t_{n+1}$ embeds any of the terms already in the sequence. If that is the case, potential non-termination is detected and the computation is stopped. Otherwise, $t_{n+1}$ is safely added to the branch and the computation proceeds.

The key to successfully debugging complex applications is to restrict the displayed information to sensitive parts of the tree. In \Glints~it is possible to tune the information displayed by the explorer by using enriched views and reporting facilities as follows.

\vspace{-.15cm}

\begin{figure}[t]
	\centering
	\includegraphics[width=\linewidth]{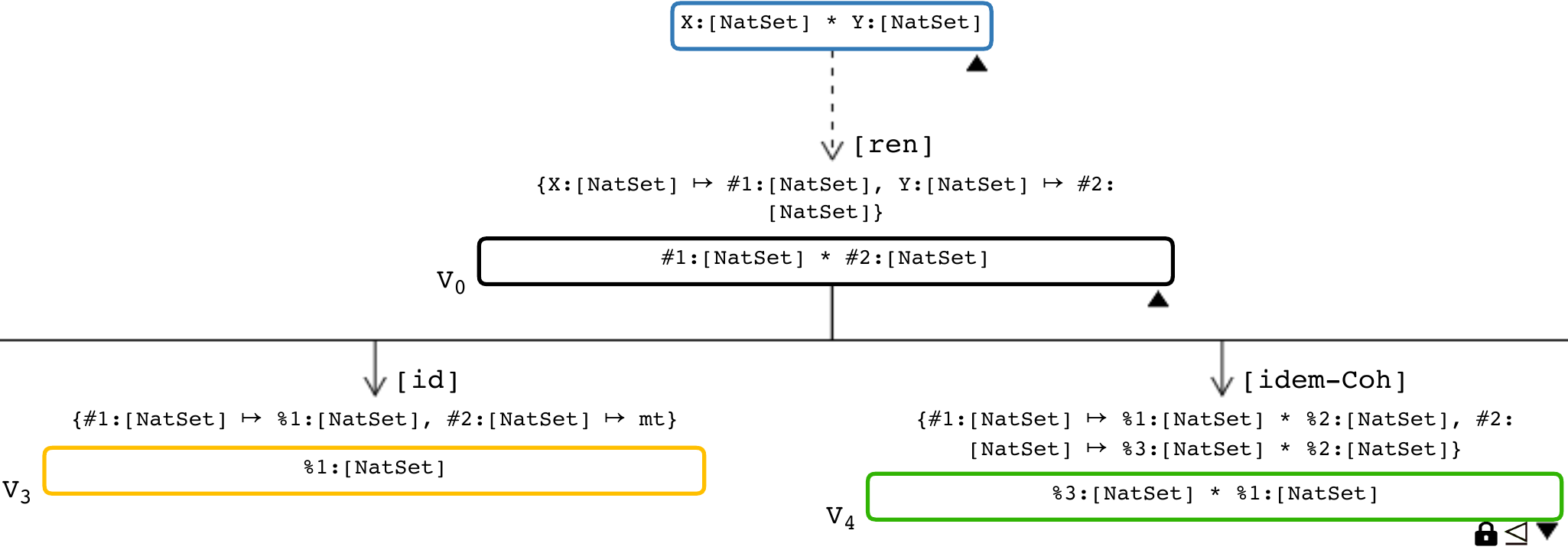}
	\caption{Enriched view showing equational unifiers for the original exclusive-or theory (fragment).}\label{fig:instrumented-tree}
\end{figure}

\paragraph{\textbf{\textit{Enriched views.}}}
\Glints~supports two distinct views, namely the standard view and the instrumented view. The standard view (which is the default mode of \Glints) focuses on the narrowing steps, whereas the instrumented view completes the picture with all the internal reduction steps that are performed up to reaching the canonical form of each variant. That is, the instrumented view reaps every single application of an equation, algebraic axiom, or built-in operation. This view is enabled by pressing the button \textit{Show normalization}. The options to show/hide the equation labels and to show/hide the unifiers that enable each narrowing step of the tree (restricted to the variables of the term, as shown in Figure \ref{fig:instrumented-tree}) are also available by two corresponding buttons.

\vspace{-.15cm}

\paragraph{\textbf{\textit{Comparing and exporting.}}}
Given the currently deployed narrowing tree, the complete list of computed variants can be shown and exported by clicking the option \textit{Export variants}. In order to easily discern the differences between two variants, a \textit{Compare variants} button is also provided that confronts two variant nodes (selected by just two consecutive clicks) in a new window where they are displayed next to each other, one on the left half of the window and the other one on the right half. \Glints~can export both the entire narrowing tree or any of its branches in two different formats, namely as an object in JSON format and as a term in Maude's meta-level representation, both of which are suitable for automated processing. This allows other tools that use \Glints~for narrowing execution to implement their own analysis on the trees delivered by \Glints. The meta-representation of terms can be visually displayed, which is particularly useful for the analysis of object-oriented computations where object attributes can only be unambiguously visualized in the meta-level (desugared) terms.

A starting guide that contains a complete description of all of the settings and detailed sessions can be found at: \url{http://safe-tools.dsic.upv.es/glints/download/quickstart.pdf}.

\vspace{-.9cm}

\section{Implementation}\label{sec:imple}

In this section, we discuss some relevant implementation details of the variant explorer \Glints.
 
\vspace{-0,25cm}
 
\subsection{Architecture of \Glints~}
\begin{wrapfigure}{r}{0.35\textwidth}
\centering
\vspace{-1cm}
\includegraphics[scale=.55]{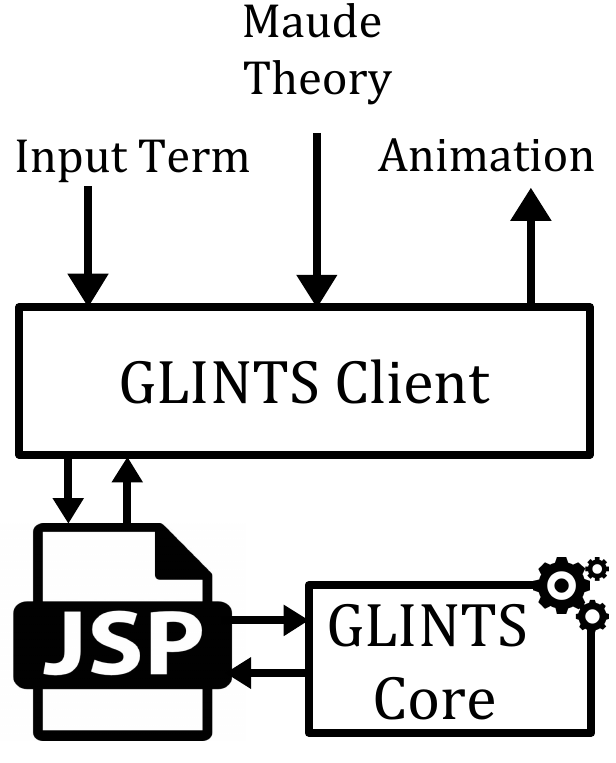}
\caption{Architecture of \Glints.}\label{fig:architecture}
\vspace{-1cm}
\end{wrapfigure} 

\Glints~has the classical architecture of a web application, which consists of two main components (the front-end and the back-end), as depicted in Figure~\ref{fig:architecture}. The two components are connected via a JSP-based layer that is implemented in Java (450 lines of Java source code). The front-end (or presentation layer) consists of 3K lines of Javascript, HTML5, and CSS source code, and provides \Glints~with an intuitive Web user interface. The back-end (or core engine) supports \Glints~services and consists of 200 function definitions (2K lines of Maude source code).
 
\subsection{Extending Maude's variant meta-operations}
One of the main challenges in the implementation of a trace-based Maude tool such as \Glints~is to make explicit the concrete sequence of internal term transformations occurring in a specific Maude computation, which is generally hidden and inaccessible within Maude's rewriting and narrowing machineries. For the case of variant narrowing computation traces, the basic information that is necessary to visually deploy the variant narrowing trees can be essentially obtained by invoking the \texttt{metaGetVariant} and \texttt{metaGetIrredundantVariant} meta-operations. That is the only way to retrieve the precise information that makes the structure of the tree explicit. Specifically, what Maude outputs is the following (in this order): (i) the computed variant term, (ii) the computed variant substitution, (iii) the largest index $n$ of any fresh variable appearing in the solutions, (iv) the identifier of the parent variant, and (v) a boolean flag that indicates whether or not there are more variants in the current tree level.

However, for the sake of efficiency, other relevant information that is key for variant narrowing debugging and understanding is not disclosed by Maude, either at the meta-level (as returned by the \texttt{metaGetVariant} and \texttt{metaGetIrredundantVariant} operations themselves) or at the source-level (as delivered in raw text format by the standard Maude interactive debugger, which furthermore cannot be manipulated as a meta-level expression by Maude). To provide the user with a deeper and more agile debugging experience, we have enriched the highly efficient developer version of Maude that we implemented in previous work, Mau-Dev\footnote{Mau-Dev has been developed under the GPLv2 license (which is the one enforced by Maude) and is fully compatible with Maude while preserving the efficiency of all standard (meta-level) operations and commands.}~\cite{ABFS16-tplp,maudev}, with two new meta-operations, namely {\tt metaGetVariantsExt} and {\tt metaGetIrredundantVariantExt}, which have been implemented in C++. By doing this, besides piecing everything together and giving a graphical reconstruction of the variant narrowing tree, \Glints\ also distills the equations, axioms, and built-in operators applied in (simplification and) narrowing steps, together with the equational unifier that enables each step.

\begin{table}[t]
	\centering
		{\setlength{\tabcolsep}{0em}
		\begin{tabular}{|c|r|r|r|r|r|}
			\cline{1-6} 
			&~~~Number~of~~~
			&\multicolumn{2}{c|}{\tt~~~~metaGetVariant~~~~~}
			&\multicolumn{2}{c|}{\tt~~~metaGetVariantExt~~~}\\
			\cline{3-6}
			&\multicolumn{1}{c|}{variants} 
			&\multicolumn{1}{c|}{~~~~~size~(kB)~~~~~}
			&\multicolumn{1}{c|}{time~(s)}
			&\multicolumn{1}{c|}{~~~~~size~(kB)~~~~~~}
			&\multicolumn{1}{c|}{time~(s)}\\
			\cline{1-6}
			\multirow{3}{*}{~~Exclusive-or~~} 
			&40~       &7.37~    		&2.49~      &12.34~     	&2.48~~\\ 
			&45~       &8.81~    		&24.82~     &14.42~     	&24.56~~\\
			&50~       &10.37~   		&302.18~    &16.62~     	&299.29~~\\ 
			\cline{1-6}
			\multirow{3}{*}{Fibonacci} 
			&40~       &520.23~		   	&3.51~      &1,417.26~    	&3.59~~\\ 
			&45~       &2,198.07~ 		&20.52~     &5,151.39~    	&20.94~~\\ 
			&50~       &5,751.55~ 		&406.59~    &15,675.13~   	&415.14~~\\ 
			\cline{1-6}
			\multirow{3}{*}{Flip-graph} 
			&500~      &4,804.66~  		&3.05~      &7,259.92~    	&3.09~~\\ 
			&1,000~     &19,520.91~		&30.33~     &29,387.01~    	&30.93~~\\ 
			&2,000~     &80,372.41~		&360.29~    &120,769.01~		&361.54~~\\ 
			\cline{1-6}
			\multirow{3}{*}{Parser} 
			&2,500~     &1,961.51~   		&3.91~      &3,067.46~ 	 	&3.92~~\\ 
			&5,000~     &5,027.82~  		&16.88~     &7,238.53~ 		&17.37~~\\ 
			&10,000~    &13,178.03~		&81.64~    &17,598.87~ 		&81.99~~\\ 
			\cline{1-6}
		\end{tabular}
		}
\vspace{-.15cm}
\caption{Execution results of the {\tt metaGetVariant} and {\tt metaGetVariantExt} operations.}\label{fig:performance}
\vspace{-.3cm}
\end{table}

Table~\ref{fig:performance} provides some figures regarding the execution of the new {\tt metaGetVariantExt} operation in comparison with the standard {\tt metaGetVariant} operation. We have tested both implementations on a 3.3GHz Intel Xeon E5-1660 with 64 GB of RAM by generating a number of variants for a collection of Maude programs that are all available at the \Glints~website: {\it Exclusive-or}, the classical specification of the boolean XOR; {\it Fibonacci}, a Maude specification that computes the Fibonacci sequence \cite{maude-manual}; {\it Flip-graph}, a variant of the classical {\it flip} function for binary graphs (instead of trees) taken from \cite{ACEM17}; and {\it Parser}, a generic parser for languages generated by simple, right regular grammars also from \cite{ACEM17}. Specifically, for each Maude program, we have asked \Glints\ to compute three different numbers of variants, which takes from a few seconds to a few minutes to generate. We have measured the {\tt metaGetVariant} invocations on a statically compiled version of the last alpha release of Maude (alpha 111a), whereas the {\tt metaGetVariantExt} invocations have been benchmarked on a Mau-Dev executable that is based on the same alpha version. 

The two {\tt size} columns correspond to the size (in kilobytes) of the generated narrowing tree (up to the requested variant), whereas the two {\tt time} columns show the average of five different measures of the computation time (in seconds). As our experiments show, the incurred overheads w.r.t. the original meta-operation are almost negligible. Note that even for extremely huge narrowing trees, the amount of data handled is much higher w.r.t.\ the original meta-operation (with an average increasement factor of $1.8$) yet the execution time is practically identical. Actually, some executions are even faster in the extended version (e.g., computing the fiftieth variant of the exclusive-or example), which can be explained by the known side-effects of Maude's garbage collector and cache memory hits and misses. Further details and runnable code are available at: \linebreak \url{http://safe-tools.dsic.upv.es/glints/pages/experiments.jsp}

\vspace{-0.35cm}

\section{Concluding remarks}\label{sec:related}

Visualization of program executions has received much attention for program debugging, optimization, profiling, and understanding in symbolic execution frameworks such as (Concurrent) Constraint Logic Programming \cite{DHM06}. However, with the exception of \Glints, no such visualization tool exists for variant narrowing computations in Maude, let alone one with the capability to reason about equational properties such as embedding and closedness modulo axioms and the finite variant property.

Besides the applications outlined in this article, further applications could benefit from variant generation in \Glints. Actually, an important number of applications (and tools) are currently based on variant generation: for instance, the protocol analyzers Maude-NPA \cite{EMM09} and Tamarin~\cite{MSCB13}, proofs of coherence and local confluence~\cite{DM12}, termination provers~\cite{DLM09}, variant-based satisfiability checkers \cite{Meseguer15}, the partial evaluator {\sf Victoria} \cite{ACEM17}, and different applications of symbolic reachability analyses \cite{BEM13}. As an application example, protocol analysis tools that rely on variant computation could identify all of the intermediate variant states that are associated to a concrete protocol state and how one is generated from the other (which is convoluted in the output provided by Maude), thereby allowing deep optimizations to cut down the search space. Indeed, many protocol analysis tools suffer from huge memory problems due to complex equational theories that generate lots of variants. 
 
As future work, we plan to address several extensions of \Glints, such as computing constructor variants~\cite{Meseguer15} and irredundant variants~\cite{maude-manual}, and supporting irreducibility constraints~\cite{EEKL+12}.

\vspace{-0,35cm}


\begin{thebibliography}{}	
	\bibitem[\protect\citeauthoryear{Alpuente, Ballis, Frechina, and
		Sapi{\~n}a}{Alpuente et~al\mbox{.}}{2016}]{ABFS16-tplp}
	{\sc Alpuente, M.}, {\sc Ballis, D.}, {\sc Frechina, F.}, {\sc and} {\sc
		Sapi{\~n}a, J.} 2016.
	\newblock {Assertion-based Analysis via Slicing with ABETS}.
	\newblock {\em Theory and Practice of Logic Programming\/}~{\em 16,\/}~5--6,
	515--532.
	
	\bibitem[\protect\citeauthoryear{Alpuente, Cuenca-Ortega, Escobar, and
		Meseguer}{Alpuente et~al\mbox{.}}{2017}]{ACEM17}
	{\sc Alpuente, M.}, {\sc Cuenca-Ortega, A.}, {\sc Escobar, S.}, {\sc and} {\sc
		Meseguer, J.} 2017.
	\newblock {Partial Evaluation of Order-sorted Equational Programs modulo
		Axioms}.
	\newblock In {\em {Proc. of the 26th Int'l Symposium on Logic-Based Program
			Synthesis and Transformation (LOPSTR 2016)}}. LNCS. {Springer}.
	\newblock To appear 2017.
	
	\bibitem[\protect\citeauthoryear{Alpuente, Falaschi, Moreno, and
		Vidal}{Alpuente et~al\mbox{.}}{1997}]{AFMV97}
	{\sc Alpuente, M.}, {\sc Falaschi, M.}, {\sc Moreno, G.}, {\sc and} {\sc Vidal,
		G.} 1997.
	\newblock {Safe Folding/Unfolding with Conditional Narrowing}.
	\newblock In {\em {Proc. of the 6th Int'l Joint Conference on Algebraic and
			Logic Programming (ALP 1997)}}. LNCS, vol. 1298. {Springer}, 1--15.
	
	\bibitem[\protect\citeauthoryear{Alpuente, Falaschi, and Vidal}{Alpuente
		et~al\mbox{.}}{1998a}]{AFV98b}
	{\sc Alpuente, M.}, {\sc Falaschi, M.}, {\sc and} {\sc Vidal, G.} 1998a.
	\newblock {A Unifying View of Functional and Logic Program Specialization}.
	\newblock {\em {ACM Computing Surveys}\/}~{\em 30,\/}~3es, 9es.
	
	\bibitem[\protect\citeauthoryear{Alpuente, Falaschi, and Vidal}{Alpuente
		et~al\mbox{.}}{1998b}]{AFV98}
	{\sc Alpuente, M.}, {\sc Falaschi, M.}, {\sc and} {\sc Vidal, G.} 1998b.
	\newblock {Partial Evaluation of Functional Logic Programs}.
	\newblock {\em {ACM Transactions on Programming Languages and Systems}\/}~{\em
		20,\/}~4, 768--844.
	
	\bibitem[\protect\citeauthoryear{Bae, Escobar, and Meseguer}{Bae
		et~al\mbox{.}}{2013}]{BEM13}
	{\sc Bae, K.}, {\sc Escobar, S.}, {\sc and} {\sc Meseguer, J.} 2013.
	\newblock {Abstract Logical Model Checking of Infinite-State Systems Using
		Narrowing}.
	\newblock In {\em {Proc. of the 24th Int'l Conference on Rewriting Techniques
			and Applications (RTA 2013)}}. {LIPIcs}, vol.~21. {Schloss Dagstuhl -
		Leibniz-Zentrum f{\"u}r Informatik}, 81--96.
	
	\bibitem[\protect\citeauthoryear{Bouchard, Gero, Lynch, and Narendran}{Bouchard
		et~al\mbox{.}}{2013}]{BGLN13}
	{\sc Bouchard, C.}, {\sc Gero, K.~A.}, {\sc Lynch, C.}, {\sc and} {\sc
		Narendran, P.} 2013.
	\newblock {On Forward Closure and the Finite Variant Property}.
	\newblock In {\em {Proc. of the 9th Int'l Symposium on Frontiers of Combining
			Systems (FroCos 2013)}}. LNCS, vol. 8152. {Springer}, 327--342.
	
	\bibitem[\protect\citeauthoryear{Chen and Warren}{Chen and Warren}{1996}]{CW96}
	{\sc Chen, W.} {\sc and} {\sc Warren, D.~S.} 1996.
	\newblock {Tabled Evaluation with Delaying for General Logic Programs}.
	\newblock {\em Journal of the ACM\/}~{\em 43,\/}~1, 20--74.
	
	\bibitem[\protect\citeauthoryear{Clavel, Dur{\'a}n, Eker, Escobar, Lincoln,
		Mart{\'i}-Oliet, Meseguer, and Talcott}{Clavel
		et~al\mbox{.}}{2016}]{maude-manual}
	{\sc Clavel, M.}, {\sc Dur{\'a}n, F.}, {\sc Eker, S.}, {\sc Escobar, S.}, {\sc
		Lincoln, P.}, {\sc Mart{\'i}-Oliet, N.}, {\sc Meseguer, J.}, {\sc and} {\sc
		Talcott, C.} 2016.
	\newblock {Maude Manual (Version 2.7.1)}.
	\newblock Tech. rep., {SRI Int'l Computer Science Lab.}
	
	\bibitem[\protect\citeauthoryear{Comon-Lundh and Delaune}{Comon-Lundh and
		Delaune}{2005}]{CD05}
	{\sc Comon-Lundh, H.} {\sc and} {\sc Delaune, S.} 2005.
	\newblock {The Finite Variant Property: How to Get Rid of Some Algebraic
		Properties}.
	\newblock In {\em {Proc. of the 16th Int'l Conference on Rewriting Techniques
			and Applications (RTA 2005)}}. LNCS, vol. 3467. {Springer}, 294--307.
	
	\bibitem[\protect\citeauthoryear{Deransart, Hermenegildo, and
		Maluszynski}{Deransart et~al\mbox{.}}{2006}]{DHM06}
	{\sc Deransart, P.}, {\sc Hermenegildo, M.~V.}, {\sc and} {\sc Maluszynski, J.}
	2006.
	\newblock {\em {Analysis and Visualization Tools for Constraint Programming:
			Constraint Debugging}}. LNCS, vol. 1870.
	\newblock {Springer}.
	
	\bibitem[\protect\citeauthoryear{Dur{\'a}n, Eker, Escobar, Mart{\'i}-Oliet,
		Meseguer, and Talcott}{Dur{\'a}n et~al\mbox{.}}{2016}]{DEEM+16}
	{\sc Dur{\'a}n, F.}, {\sc Eker, S.}, {\sc Escobar, S.}, {\sc Mart{\'i}-Oliet,
		N.}, {\sc Meseguer, J.}, {\sc and} {\sc Talcott, C.} 2016.
	\newblock {Built-in Variant Generation and Unification, and their Applications
		in Maude 2.7}.
	\newblock In {\em {Proc. of the 8th Int'l Joint Conference on Automated
			Reasoning (IJCAR 2016)}}. LNCS, vol. 9706. {Springer}, 183--192.
	
	\bibitem[\protect\citeauthoryear{Dur{\'a}n, Lucas, and Meseguer}{Dur{\'a}n
		et~al\mbox{.}}{2009}]{DLM09}
	{\sc Dur{\'a}n, F.}, {\sc Lucas, S.}, {\sc and} {\sc Meseguer, J.} 2009.
	\newblock {Termination Modulo Combinations of Equational Theories}.
	\newblock In {\em {Proc. of the 7th Int'l Symposium on Frontiers of Combining
			Systems (FroCos 2009)}}. LNCS, vol. 5749. {Springer}, 246--262.
	
	\bibitem[\protect\citeauthoryear{Dur{\'{a}n} and Meseguer}{Dur{\'{a}n} and
		Meseguer}{2012}]{DM12}
	{\sc Dur{\'{a}n}, F.} {\sc and} {\sc Meseguer, J.} 2012.
	\newblock {On the Church-Rosser and Coherence Properties of Conditional
		Order-sorted Rewrite Theories}.
	\newblock {\em The Journal of Logic and Algebraic Programming\/}~{\em
		81,\/}~7--8, 816--850.
	
	\bibitem[\protect\citeauthoryear{Erbatur, Escobar, Kapur, Liu, Lynch, Meadows,
		Meseguer, Narendran, Santiago, and Sasse}{Erbatur
		et~al\mbox{.}}{2012}]{EEKL+12}
	{\sc Erbatur, S.}, {\sc Escobar, S.}, {\sc Kapur, D.}, {\sc Liu, Z.}, {\sc
		Lynch, C.}, {\sc Meadows, C.}, {\sc Meseguer, J.}, {\sc Narendran, P.}, {\sc
		Santiago, S.}, {\sc and} {\sc Sasse, R.} 2012.
	\newblock {Effective Symbolic Protocol Analysis via Equational Irreducibility
		Conditions}.
	\newblock In {\em {Proc. of the 17th European Symposium on Research in Computer
			Security (ESORICS 2012)}}. LNCS, vol. 7459. {Springer}, 73--90.
	
	\bibitem[\protect\citeauthoryear{Escobar, Meadows, and Meseguer}{Escobar
		et~al\mbox{.}}{2009}]{EMM09}
	{\sc Escobar, S.}, {\sc Meadows, C.}, {\sc and} {\sc Meseguer, J.} 2009.
	\newblock {Maude-NPA: Cryptographic Protocol Analysis Modulo Equational
		Properties}.
	\newblock In {\em {Foundations of Security Analysis and Design V (FOSAD
			2007/2008/2009 Tutorial Lectures)}}. LNCS, vol. 5705. {Springer}, 1--50.
	
	\bibitem[\protect\citeauthoryear{Escobar, Sasse, and Meseguer}{Escobar
		et~al\mbox{.}}{2012}]{ESM12}
	{\sc Escobar, S.}, {\sc Sasse, R.}, {\sc and} {\sc Meseguer, J.} 2012.
	\newblock {Folding Variant Narrowing and Optimal Variant Termination}.
	\newblock {\em The Journal of Logic and Algebraic Programming\/}~{\em
		81,\/}~7--8, 898--928.
	
	\bibitem[\protect\citeauthoryear{Hanus}{Hanus}{2013}]{Hanus13}
	{\sc Hanus, M.} 2013.
	\newblock {Functional Logic Programming: From Theory to {Curry}}.
	\newblock In {\em {Programming Logics. Essays in Memory of Harald Ganzinger}}.
	LNCS, vol. 7797. {Springer}, 123--168.
	
	\bibitem[\protect\citeauthoryear{Leuschel}{Leuschel}{2002}]{Leuschel02}
	{\sc Leuschel, M.} 2002.
	\newblock {Homeomorphic Embedding for Online Termination of Symbolic Methods}.
	\newblock In {\em {The Essence of Computation. Essays Dedicated to Neil D.
			Jones on the Occasion of his 60th Birthday}}. LNCS, vol. 2566. {Springer},
	379--403.
	
	\bibitem[\protect\citeauthoryear{??}{Mau-Dev}{2016}]{maudev}
	Mau-Dev 2016.
	\newblock {The {\sf Mau-Dev} Web site}.
	\newblock {Available at: \url{http://safe-tools.dsic.upv.es/maudev}}.
	
	\bibitem[\protect\citeauthoryear{Meier, Schmidt, Cremers, and Basin}{Meier
		et~al\mbox{.}}{2013}]{MSCB13}
	{\sc Meier, S.}, {\sc Schmidt, B.}, {\sc Cremers, C.}, {\sc and} {\sc Basin,
		D.~A.} 2013.
	\newblock {The TAMARIN Prover for the Symbolic Analysis of Security Protocols}.
	\newblock In {\em {Proc. of the 25th Int'l Conference on Computer Aided
			Verification (CAV 2013)}}. LNCS, vol. 8044. {Springer}, 696--701.
	
	\bibitem[\protect\citeauthoryear{Meseguer}{Meseguer}{1992}]{Meseguer92}
	{\sc Meseguer, J.} 1992.
	\newblock {Conditional Rewriting Logic as a Unified Model of Concurrency}.
	\newblock {\em Theoretical Computer Science\/}~{\em 96,\/}~1, 73--155.
	
	\bibitem[\protect\citeauthoryear{Meseguer}{Meseguer}{2015}]{Meseguer15}
	{\sc Meseguer, J.} 2015.
	\newblock {Variant-Based Satisfiability in Initial Algebras}.
	\newblock In {\em {Proc. of the 4th Int'l Workshop for Safety-Critical Systems
			(FTSCS 2015)}}. Communications in Computer and Information Science, vol. 596.
	{Springer}, 3--34.
	
	\bibitem[\protect\citeauthoryear{Meseguer}{Meseguer}{2006}]{Meseguer06}
	{\sc Meseguer, T.} 2006.
	\newblock {From OBJ to Maude and Beyond}.
	\newblock In {\em {Proc. of Algebra, Meaning, and Computation. Essays Dedicated
			to Joseph A. Goguen on the Occasion of His 65th Birthday}}. LNCS, vol. 4060.
	{Springer}, 252--280.
	
	\bibitem[\protect\citeauthoryear{Pettorossi and Proietti}{Pettorossi and
		Proietti}{1996}]{PP96}
	{\sc Pettorossi, A.} {\sc and} {\sc Proietti, M.} 1996.
	\newblock {A Comparative Revisitation of Some Program Transformation
		Techniques}.
	\newblock In {\em {Int'l Dagstuhl Seminar on Partial Evaluation}}. LNCS, vol.
	1110. {Springer}, 355--385.
	
	\bibitem[\protect\citeauthoryear{Yang, Escobar, Meadows, Meseguer, and
		Narendran}{Yang et~al\mbox{.}}{2011}]{YEMMN14}
	{\sc Yang, F.}, {\sc Escobar, S.}, {\sc Meadows, C.}, {\sc Meseguer, J.}, {\sc
		and} {\sc Narendran, P.} 2011.
	\newblock {Theories of Homomorphic Encryption, Unification, and the Finite
		Variant Property}.
	\newblock In {\em {Proc. of the 16th Int'l Symposium on Principles and Practice
			of Declarative Programming (PPDP 2014)}}. {ACM Press}, 123--133.	
\end{thebibliography}

\end{document}